\documentclass[a4paper,11pt]{article}
\pdfoutput=1 

\usepackage{jheppub} 

\usepackage{amsthm}
\usepackage{amsfonts}
\usepackage{mathtools}
\usepackage{xcolor}

\def\W{\mathcal{W}}

\def\a{\alpha}

\def\inf{\infty}
\def\bsp{\begin{split}}
\def\esp{\begin{split}} 
\def\N{\mathcal{N}}
\def\Z{\mathbb{Z}}

\def\r{\gamma}

\newtheorem{theorem}{Theorem}
\def\bee{\begin{equation}}
\def\eee{\end{equation}}
\def\bsp{\begin{split}}
\def\esp{\end{split}}
\def\u{\mu}
\def\w{\omega}

\title{R-Twisting, Fibered Knots, and Gauge Theories }

\author[-,+]{Shi Cheng}

\affiliation[-]{Center for Mathematics
and Interdisciplinary Sciences, Fudan University
Shanghai 200433, China}

\affiliation[+]{Shanghai Institute for Mathematics
and Interdisciplinary Sciences
Shanghai 200433, China}

\emailAdd{shicheng@simis.cn}

\abstract{
We argue that R-twisting implies the Seiberg-Witten curves of 4d Argyres-Douglas theories could be combined with the R-twisted circle to form the mapping tori of torus knots, inspired by Milnor fibration theorem. Then 3d gauge theories labeled by  mapping tori  are generated, which in the IR capture 4d BPS spectrum and mutations. This construction addresses the boundary problem of the domain wall approach in \cite{Cecotti:2011iy}. These 3d theories show a supersymmetry enhancement, and are related to 3d rank zero theories.
}

\begin{document} 

\maketitle

\section{Introduction}

It is expected that the 3d $\N=2$ gauge theories can be engineered by three-manifolds, by wrapping 6d $(2,0)$ superconformal field theories on either hyperbolic  or non-hyperbolic three-manifolds \cite{Dimofte:2011ju}. Another perspective is that many 3d gauge theories should in principle come from  4d theories, by using the domain wall construction \cite{Cecotti:2011iy,TerashimaYamazaki2011} or the circle reduction \cite{Benini:2010uu}. 

In recent years, a large class of 3d theories called rank zero theories draw attentions \cite{GangYamazaki2018,Dedushenko:2023cvd}, which inspire people to revisit the relation between 4d gauge theories and 3d gauge theories. 
One key concept is the R-twisting \cite{Cecotti:2009uf,Cecotti:2010fi} that plays a central role by connecting many concepts. This motivates us to think about how to understand R-twisting in the geometric engineering of 3d gauge theories. 

The key idea of this paper comes from the observation that the polynomials defining Milnor fibrations \cite{Milnor1968Singular} of the mapping tori are the same as the Seiberg-Witten curves of Argyres-Douglas theories. So our key idea is that the R-twisted circle given by the R-twisting should be combined with the Seiberg-Witten curves in a non-trivial way to form three-manifolds, which are mapping tori of torus knots with non-trivial monodromies. We argue that 3d gauge theories relate to 4d gauge theories by this construction.

A wonderful property of the mapping tori $\Sigma_{SW} \times_h S^1_t$ of torus knots $T^{(m,n)}$ is that they are just the torus knot complements $S^3 \setminus T^{(m,n)}$, which lead to significant simplification for considering Dehn surgeries, Kirby moves, and so on, and allow us to read off the gauge theories from the topology. In addition, the irregular punctures of the AD theories encode these torus knots. 

Mapping tori are friendly to the domain wall construction of 3d gauge theories, since the boundary issue does not exist this time. Following \cite{Cecotti:2010fi,Cecotti:2011iy}, we can consider the wall crossing of BPS quivers on the mapping tori, and everything is consistent with the monodromy. 

Another nice property of mapping tori is that they allow transverse holomorphic foliations, which lead to a larger supersymmetry $\N=4$, and coincide with the enhancement condition in \cite{Assel_2023}. Since both our method and \cite{Assel_2023} lead to this same result, we believe that the mapping tori give 3d $\N=4$ theories, and these theories should relate to the 3d rank zero theories.

The structure of this paper is as follows: In section \ref{sec:mappingtori}, we review the mapping tori and Milnor fibration. In section \ref{sec:Rtwisitng}, we reformulate R-twisting to match with Milnor fibration. In section \ref{sec:domainwall}, we combine mapping tori and M2-M5-brane configuration, and remove conceptual barriers in the domain wall construction. 
In section \ref{sec:surgery}, we switch the language to surgery, and discuss the M2-branes ending on three-manifolds and corresponding gauge theories.
In section \ref{sec:enhancement}, we discuss the supersymmetic enhancement.

\section{Mapping tori}
\label{sec:mappingtori}

Consider the elliptic bundle over a circle. When the two-manifold goes around the circle, the two-manifold experiences a homeomorphism given by an element in the mapping class group of the two-manifold. This morphism is called the monodromy. The mapping tori are defined by 
\begin{align}
M_h = \Sigma_g \times [0,1]/\sim \,,
\end{align} 
where $(x,0) \sim (h(x),1)$ and $h$ is the monodromy. The two-manifold has genus $g$, and may have punctures given by digging out some two-dimensional discs $D^2$, and these punctures are the boundaries of  the mapping tori. For example, and $\Sigma_g \times S^1$ is  a mapping tori with trivial monodromy $h=1$.

\subsection {Open book decomposition}

\begin{figure}[htp!]
        \centering
     \includegraphics[width=1in]{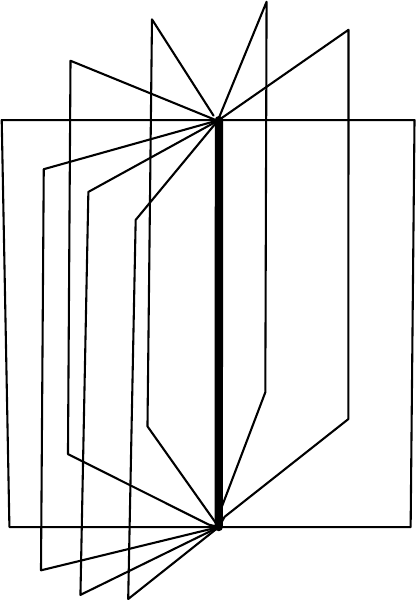}
        \caption{The thick line as the common edge is a knot, and each page is isotopic to the Seifert surface bounded by the knot. The monodromy shifts from one page to the next page. This whole book is the closed three-manifold. The most basic example is the unknot, which is a line extending to infinity.}
        \label{fig:openbook}
    \end{figure}

It is well known that any three-manifold allows a open book decomposition (OBD). The OBD is illustrated in Figure \ref{fig:openbook}, and means the three-manifold can be foliated, and all leaves have a common edge as a knot in general.

For some special knots,  whose complements $S^3 \setminus K $ are  a mapping tori, which is a beautiful connection, and  these knots are called fibered knots. The open book decomposition and mapping tori are the same structure for these knot complements, because each page is the Seifert surface,  the common edge is the fibered knot, and the base $S^1$ of the mapping tori is giving by going around the common edge. 
This implies that the fibered knot complement is a bundle:
\begin{align}
  S^3\setminus K \rightarrow S^1 \,.
\end{align}
Notice that the base $S^1$ is not lifted to the knot $K$ itself, but its meridian. 

If there is a puncture on the Seifert surface, then this puncture will trace out a torus hole $T^2 =S^1\times S^1$ in the mapping tori if going around the base $S^1$ This torus hole is just the  boundary of the knot complement $S^3 \setminus K$. 
Putting these two perspective together, we have
\begin{align}
M_h=\Sigma_{g,1} \times_h S^1  = S^3\setminus K  \,.
\end{align}
Sometimes, this fibered knot can be a link, such as Hopf link which has two punctures on the Seifert surface and the boundaries of the Hopf knot complement form a Hopf link.

The punctures on Seifert surfaces can be filled by gluing in  two-discs $D^2$, and then the fibered knot complement become closed three-manifolds. Correspondingly, in three-manifold, this filling lifts to gluing in a solid torus $D^2\times S^1$ to the knot complement, which is  the operation called  Dehn surgery. In addition, there is something called framing numbers that label the gluing maps for surgeries. If a puncture on the Seifert surface is filled by a two-disc $D^2$, then the framing number is zero. This case is called 0-surgery, and the closed three-manifold obtained is denoted by $S^3_0(K)$.

The most simple example is the unknot; then the three-manifold is $S^3 \setminus \bigcirc$. After the 0-surgery, it becomes a $S^2\times S^1$. However, we do not consider this simple case, since its related gauge theories are already known. We will focus on torus knots, which consist a large class of fibered knots.

\subsection{Seifert surface}
\begin{figure}[htp!]
        \centering
     \includegraphics[width=2in]{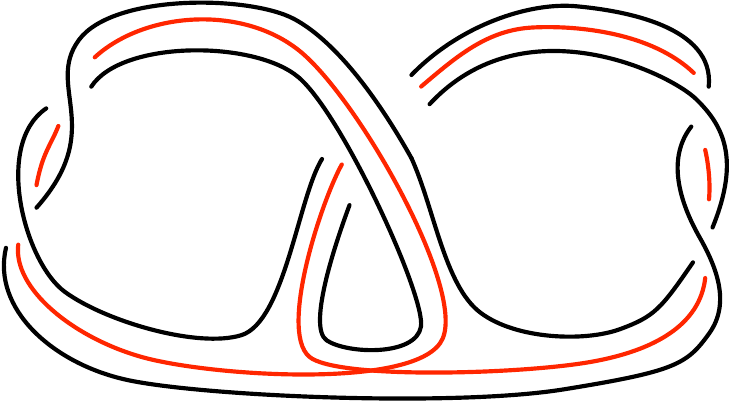} \quad \quad 
      \includegraphics[width=2in]{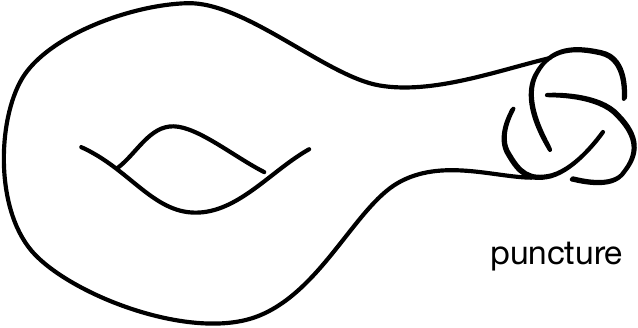}
        \caption{The band graph and Seifert surface for the trefoil.}
        \label{fig:trefoilseifert}
    \end{figure}

The knot $K$ gives rise to the Seifert surface that reflects interesting properties of the mapping tori. 
Firstly, we should know how the knot bounds the Seifert surface. There is a canonical procedure: the knot should be deformed as the band graph, and then the bounded region is the Seifert surface. For example, the trefoil leads to the brand graph consisting of two handles, as illustrated in Figure  \ref{fig:trefoilseifert}.
    The two-dimensional one-handle is a rectangle $I \times I$, and can be glued to a disc $D^2$ to form the band graph. 
    This Seifert surface can be shown to be a 2-manifold $\Sigma_{1,1}$, although it is not obvious at first sight. The red circles on the band graph are just the longitude and meridian of the $\Sigma_{1,1}$.

A surprise is that the boundary of the Seifert surface is a knot that encodes the topology of the puncture of the two-manifold. 
For the 2-manifold itself, it cannot tell if the puncture is a knot or an unknot.  When it is embodied in a higher dimensional manifold, such as the three-manifold, can the topology of the puncture be identified. Recall that this knot $K$ is just the knot of the complement $S^3 \setminus K$. The Seifert surface $\Sigma_{g,1}$ is inserted in the complement $S^3 \setminus K$ in a highly twisted pattern. When the fiber of mapping tori goes around the base circle, correspondingly this Seifert surface rotates around the knot $K$ in the knot complement. A good illustration is that the knot $K$ is the spine of a book, and each page of the book is the Seifert surface.
When we move from a page to other pages, the knot $K$ is the fixed edge under this rotation, which is a much more complicated configuration than the one in Figure \ref{fig:openbook}.
See \cite{Akbulut_2001,Rolfsen1976,Saveliev1999Lectures} for nice explanations and more details.

 Torus knots can be drawn on a torus, by  winding both longitude and meridian, and hence are denoted by winding numbers ${(m,n)}$. For instance, the trefoil $\textbf{3}_1$ is the $T^{(2,3)}$ torus knot, and Hopf link is the torus knot $T^{(2,2)}$.
For $T^{(m,n)}$, the bounded Seifert surface has genus $g={(m-1)(n-1)}/{2}$. This formula should be slightly modified when the formula does not give an integer.
We have shown the Seifert surface for trefoil in Figure \ref{fig:trefoilseifert}.
 The Seifert surfaces for generic torus knots can be obtained by repeating this pattern of adding 1-handles either horizontally or vertically; see e.g. \cite{Akbulut_2001}.

\vspace{4mm} \noindent
\textbf{Seifert matrix.}
The monodromy $h$ of the mapping tori induces a map: $h_\star: H_1(\Sigma) \rightarrow H_1(\Sigma)$.
A subtle point is that a fibered knot can provide many different Seifert surfaces, depending on how to shade the regions bounded by the knots, which are related by Reidemeister moves. 
Even for the Seifert surface in Figure \ref{fig:trefoilseifert}, one can still get many band graphs, related by handle sliding of 2-manifolds. These band graphs are actually equivalent. The canonical Seifert surface given by the band graph is the one we should select to match gauge theories, whose information is encoded in a matrix $M=(S^T)^{-1}S$ where $S$ is the Seifert matrix determined by the intersection numbers of Seifert circles on the Seifert surfaces. The $M$ here is acturally the the monodromy matrix induced by the monodromy $h$ by acting on the first homology.

To determine the Seifert matrix, we should follow the procedure: First, draw the Seifert circle $\gamma_i$ around each component (1-handle) of the band graph, and Seifert matrix is given by the linking numbers between these circles $S_{ij} =\gamma_i \circ \gamma_j^+ $ where $+$ means the circle $\gamma_j$ is slightly lifted above the band graph to resolve the possible intersection with $\gamma_i$, and hence $S_{ij}$ is not symmetric. For the torus knots, only $\r_i$ and $\r_{i+1}$ intersect at a point. This definition gives a way to transform the intersection numbers to linking numbers of circles in the  mapping tori. 

Once we have a two-manifold, the hemomorphism that we can apply is the Dehn twist $T_{\r_i}$, which is the operation that cutting the circle $\r_i$ on a tube component of the two-manifold, recombining other circles with this circle, and then gluing back the tube, following the rule: 
\begin{align}   
&\gamma_i \rightarrow \gamma_i,\, \\
&\gamma_j \rightarrow \gamma_j + (\gamma_i \circ \gamma_j^+) \gamma_i  = \r_j + \r_i \cdot S_{ij}\,,
\end{align}
which is identified with  handle slides of two-manifolds, and is also known as the Picard-Lefschetz transformation. Dehn twists change the basis $\{\r_i\}$ of the homology $H_1(M_h)$. When a circle goes around the base $S^1$, the monodromy will change it to $M_{ij}\gamma_j$.
The monodromy of the mapping tori is expressed as the product of Dehn twists in sequence:
\begin{align}
M=T_{\r_1}T_{\r_2}\cdots T_{\r_\mu}
\end{align}
where $\mu =2g$, and these $\{\r_i\}$ can be drawn on the Seifert surface in sequence:
\begin{align} \label{Tcircle}
\bsp  \includegraphics[width=2in]{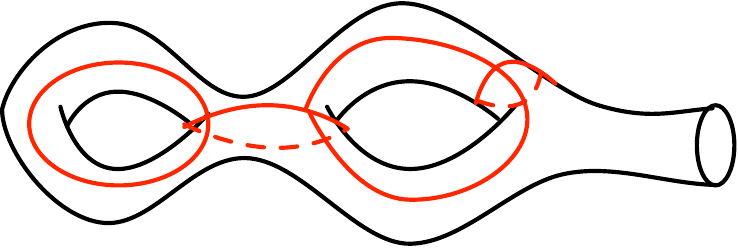}
\esp
\end{align}
Since the Dehn twist is expressed as an element in the mapping class group of the two-manifold, the monodromy can be expressed a matrix.

\subsection{Milnor fibration}
The above is  the topological aspect of mapping tori. There is an algebraic realization of the open book decomposition. Torus knots are algebraic knots described by polynomials. This is very useful because the Seiberg-Witten curves coincide with the algebraic polynomials of torus knots. This observation is the starting point of this work. 
 In this subsection, we review the Milnor fibration, which algebraically realize mapping tori of fibered knots.

\begin{theorem}[John Milnor\cite{Milnor1968Singular}]
Assume a polynomial function $f: \mathbb{C}^2 \rightarrow \mathbb{C}$ with $f(0,0)=0$ and $(0,0)$ is an isolated critical point.
Then there is an $\epsilon >0$, such that $S^3$ is the unit sphere of radius $\epsilon$, $K=f^{-1}(0) \cap S^3$, and the fibration $\varphi: S^3\setminus K \rightarrow S^1$, where
\begin{align} \label{varphi}
    \varphi (x,y) =\frac{ f(x,y)}{|f(x,y)|}
\end{align}
\end{theorem}
This theorem states that for algebraic knots, such as torus knots, the knot complement is foliated and  is a mapping tori. The Seifert surface is given by Milnor fiber $\Sigma=\varphi^{-1}(t)$ for any $t\in S^1$. In addition, there is a hiding automorphism acting on complex coordinates. For instance, the polynomials for the torus knots are given by $f(x,y)=x^p+y^q$ and the automorphism is
\begin{align}\label{milnormonodromy}
h_t: (x,y) \rightarrow ( e^{i t/p} x, e^{i t /q} y)\,,
\end{align}
which algebraically defines the monodromy of mapping tori, and ensures that if going around the base $S^1$, the polynomial is identified. The base is parameterized by $t \in [0, 2\pi]=S^1$, and the fiber is given by $\varphi^{-1}(t) = \Sigma_{g,n}$, where $n=0$ or $1$, depending on the knot. In addition, there is a Milnor number defined by $\mu=(p-1)(q-1)$, counting the number of independent circles $\gamma_i$ in the homology of the Seifert surface.
\begin{figure}
    \centering
\includegraphics[width=2in]{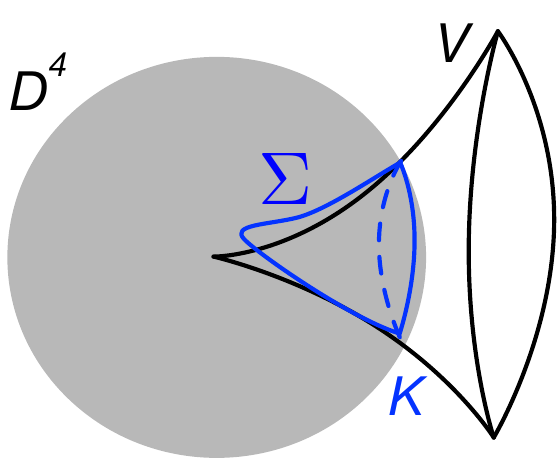}
    \caption{The Milnor fiber as the Seifert surface is defined by shifting away from the singularity $f(x,y)=0$ and taking all values in the complex plane $f(x,y) \in \mathbb{C}\setminus \{0\}  $. Hence, the intersection as a knot sweeps a surface on the boundary of $D^4$.}
\label{fig:Milnorknot}
\end{figure}

The Milnor fibration can be illustrated by Figure \ref{fig:Milnorknot}.
To be more precise, the  torus knot $T^{(p,q)}$ is the intersection of the hypersurface $V$ defined by $V=f^{-1}(0)$ and $ f(x,y) = x^p +y^q$
with the three-sphere $S^3=\{ (x,y) ~|~ |x|^2+|y|^2 =\epsilon \}$ which is the boundary of the four-dimensional ball $D^4=\{ (x,y) ~|~ |x|^2+|y|^2 \leq \epsilon \}$. The intersection $V \cap S^3=T^{(p,q)}$ bounds the Seifert surface  as the Milnor fiber.
The Seifert circles $\r_i$ are in the homology $H_1(\Sigma_{g,n} )$  and extend to discs $D_i^2$ in the bulk of of the four-ball $D^4$ and $\gamma_i =\partial D_i^2$. 
The real parameter $\epsilon$ is the length of the circle $\gamma_i$, and
each circle $\gamma_i$ shrinks to a point when $\epsilon \rightarrow 0$. In addition, the knot $K$ can be represented in terms of these $\gamma_i$. In Figure \ref{fig:trefoilseifert}, we give an example to illustrate the Seifert surface and the Seifert circles $\r_i$. These circles are called vanishing circles $\{\gamma_i\}$ as they shrink at the singularity $\epsilon =0$, and they form the wedge sum $\r_1\vee \r_2 \vee\cdots \vee \r_\mu$ where $\vee$ means the adjoint circles are connected at one point. The number of these circles is just the Milnor number $\mu$. 
It is possible to consider other algebraic knots except torus knots. For more details, see Milnor's book \cite{Milnor1968Singular}.

The Milnor fibration is consistent with M2-M5 brane configuration and Seiberg-Witten curves of Argyres-Douglas (AD) theories, which is the main tool that we can rely on to analyze gauge theories.

\section{R-twisting}\label{sec:Rtwisitng}
In this section, we relate the Milnor fibration to the R-twisting 
introduced in \cite{Cecotti:2009uf,Cecotti:2010fi}. The R-twisting is analogous to the topological twist, which is defined to identify the  $U(1)_r$ R-symmetry with a rotation symmetry $U(1)_t$ of the three-manifold. 
We conjecture that this identification should be slightly scaled to match the gauge theories with the Milnor fibration   of $(m,n)$-torus knots. This scaling is 
\begin{align}
t= \alpha \theta \,, ~~~\a= \frac{mn}{m+n} \,,
\end{align}
where $(\theta, t) $ are coordinates of $(S^1_t, S^1_\theta)$ as R-twisted circle and R-symmetry circle, respectively. The $(m,n)$-torus knots appear at the irregular punctures of 4d AD theories. This scaling can be viewed as assigning a R-charge $\a$ to the R-twisted circle $S^1_t$, which is previously not proposed.

\subsection{R-twisted circle}

Recall that the 4d spacetime can be separated as $\mathbb{R}^3 \times S_t^1$, and the 4d theories are engineered by wrapping M5-branes on Seiberg-Witten curves $\Sigma_{SW}$. In the infrared (IR), BPS states can be described by the Seiberg-Witten curves. If we relocate this circle $S^1_t$ of the 4d spacetime to the internal space and combine it with the Seiberg-Witten curve, then roughly we get a 3d theories on $\mathbb{R}^3$, labeled by $\Sigma_{SW} \times S_t^1$.
 In the geometric engineering of 4d theories, the $U(1)_r$ R-symmetry is realized as the rotation symmetry of a plane $\mathbb{R}^2$ in the normal direction $\mathbb{R}^5$ of the Seiferg-Witten curve.  The circle $S^1_\theta$ for this plane is in the internal space, and should not be confused with the R-twisted circle $S^1_t$ in the spacetime. 
 
In \cite{Cecotti:2010fi}, this circle $S_t^1$ is R-twisted with the $S^1_\theta$ of the R-symmetry $U(1)_r$ of 4d theories, in order to preserve the right number of supersymmetries of the two-dimensional sigma model after further dimensional reduction \cite{Cecotti:2009uf}. From the perspective of the 6d SCFTs, the 3d theories $T[\Sigma_{SW} \times S_t^1]$ is obtained by two steps of compactification
\begin{align}
    6d \xrightarrow{~\Sigma_{SW}} 4d~T[\Sigma_{SW}] \xrightarrow{~S_t^1} 3d
\end{align}
and  the three-manifold at here is a trivial mapping tori with a trivial monodromy.

Let us use polynomials to explain how the R-twisting is defined.
For the $(A_{m-1}, A_{n-1})$ AD theories, the Seiberg-Witten curve is given by 
\begin{align}
f(x,y)=x^m + y^n  \,.
\end{align}
The R-charge of the coordinates are $[x]=\frac{n}{m+n}$ and $[y]=\frac{m}{m+n}$, and hence when we go around the circle $S^1_\theta$,  coordinates are transformed as 
\begin{align}\label{monodromyRsym}
(x,y) \mapsto (x',y')=(e^{2 \pi i [x] \theta} x ,\, e^{2\pi i [y] \theta } y ) \,,
\end{align} 
where $\theta \in [0,1]/\sim$ is the angular coordinate $S^1_\theta$. 
The Seiberg-Witten curve  is then shifted by 
\begin{align}\label{Rphase}
f(x,y) \mapsto  f'(x,y)=e^{2\pi i \theta  \cdot \frac{mn}{m+n} } \left( x^m +y^n\right) = e^{2\pi  i \cdot \a \theta }  \cdot f(x,y)  \,,
\end{align}
where a phase is produced by the R-symmetry.

We think the R-twisting could lead us to more complicated three-manifolds, in particular non-trivial mapping tori. There is a subtle problem in front of us. To state and solve it,
let us compare with mapping tori. When we go around the base of the Milnor fibration, the Seifert surface goes back to the original one, and no additional phase is caused by the monodromy $h$. Then how to deal with the phase in \eqref{Rphase}? 
The solution is simple:
If we rewrite the action of the $U(1)_r$ R-symmetry by a scale, then we get the standard automorphism \eqref{milnormonodromy} for Milnor fibration:
\begin{equation}
( x, y) \mapsto (x',y')= ( e^{2\pi i  \cdot \frac{t}{m}} x, e^{2\pi i  \cdot \frac{t}{n}} y) \,,
\end{equation}
where $t= \a \theta$ and $\alpha:={ \frac{mn}{m+n}  }$.
If we go around the base $S^1_t$, the phase in front of $f(x,y)$ should be one, such that the Milnor fiber can be glued to form a mapping tori under monodromy. Hence the base $S^1_t$ have to take value between $[0,1]/\sim$. This is  ensured by the definition $\frac{f(x,y)}{|f(x,y)|}= e^{2 \pi i t} \in S_t^1$.

Now we have conjectured that the R-twisted circle $S^1_t$ should be related to the R-symmetry circle $S^1_\theta$ by a scale $\a$. This means that these two circles are not been winded at the same speed. It is known that the monodromy \eqref{monodromyRsym} of the R-symmetry have an order of $m+n$, which means if going around the $S^1_\theta$ by $m+n$ times, then everything goes to the origin and its action on the complex coordinates $(x,y)$ is trivial, while the order of the monodromy of the mapping tori is $mn$, and if going around it $mn$ times, the monodromy $h^{mn}=1$. This disagreement can be explained by the scaling, since 
 the lengths of multiple winding for both are equal $ (m+n) \times 1 =mn \times \frac{m+n}{mn}$. This fills the gap that in physical papers the order of monodromy is $m+n$, while the order of monodromy in Milnor fibration of mapping tori is $mn$, just because these two are viewed from different circles: The former is from the $S_\theta^1$ of $U(1)_r$, and the latter is from the base $S^1_t$ of the mapping tori. 
We illustrate this scaling by the following graph:
\begin{equation}\begin{split}
    \includegraphics[width=2in]{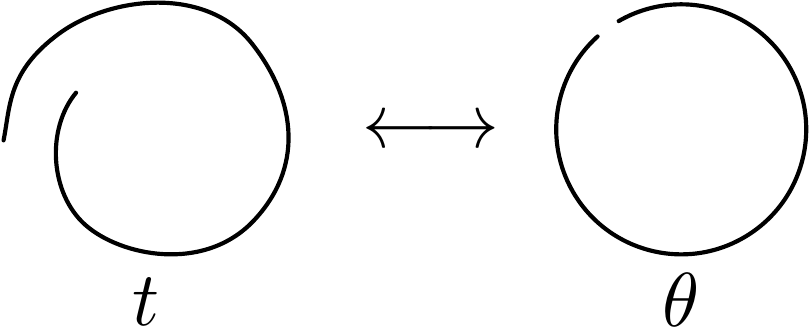} \end{split}
\end{equation}
 The scaling for the R-twisting is analogous to the $\Omega$-deformation: when going around the M-theory circle, the spacetime coordinates are only shifted by a small angle by $q= e^{2\pi i \,R \theta}$, where $R$ is the scaling.

Imposing this scaling has many benefits and should be necessary. Firstly, the trivial product $\Sigma_{SW} \times S^1$ is lifted to a non-trivial mapping tori of torus knots.  Many nice properties of mapping tori can thus be used to study gauge theories, including the transverse holomorphic foliations and so on, which are responsible for supersymmetries and we will discuss in the last section.

\section{Domain wall construction}
\label{sec:domainwall}
In the last section, we have shown the mathematical structures. From this section, we connect them to gauge theories and geometric engineering.       

\subsection{M2-branes and M5-branes}
The Milnor fibration of the mapping tori contains rich geometric structures to carry both M2-branes and M5-branes.
In string theory, the Seifert surface as the Milnor fiber plays the role of Seiberg-Witten curves of 4d theories and is wrapped by M5-branes. The shrinking circles  $\{\r_i\}$ in the Milnor fiber are the boundaries of M2-branes ending on the M5-branes.
These M2-branes are responsible for BPS states. Since the M5-brane wraps the mapping tori, in principle we can determine the details of these M2-branes, including locations and charges. Let us use a picture to illustrate this brane configuration:
\begin{align} \bsp
\includegraphics[width=1.5in]{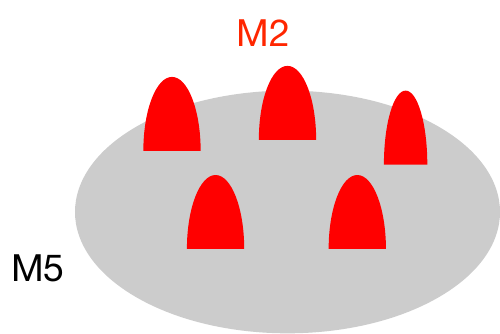}  \esp
\end{align}
This whole configuration determines the 3d theory. The M2-branes look like some fruits distributed on the trees of three-manifolds.

There is a corresponding construction in IIB string theory \cite{Alim:2011ae,Alim:2011kw}, where
the Calabi-Yau three-folds is used to engineer AD theories, and vanishing circles correspond to vanishing spheres wrapped by D3-branes. These vanishing spheres intersect with each others, and give rise to BPS quivers.
In this note, we do not consider this construction.

\subsection{Walls and BPS states}
The mapping tori provides a domain wall construction. In \cite{Cecotti:2011iy,Dimofte:2013lba}, the domain wall construction of 3d theories are considered, which requires to foliate the three-manifold into slices, and each slice is the Seiberg-Witten curve for the 4d theory, and the 3d theory lives on the wall between two 4d theories. It is possible to have many walls in the three-manifolds, and then the whole three-manifold $\Sigma_{SW} \times I$ determine the 3d theory. 
In the case of the mapping tori, the interval $I$ is replaced by a circle, and hence there is no boundary problem. We can use an annulus to illustrate the mapping tori as a domain wall: 
\begin{align} \includegraphics[width=1.5in]{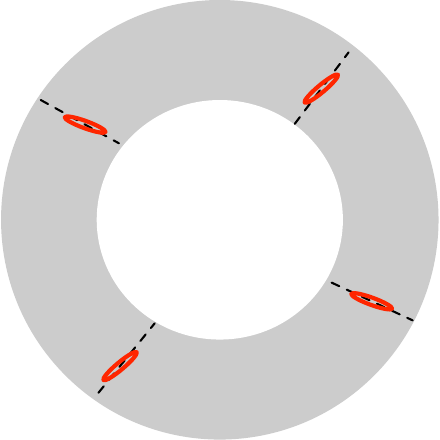}
\end{align}
where the dashed lines are the walls, on which the 3d BPS states should live. These 3d BPS states are actually selected from the spectrum of 4d BPS states. This is one main problem that we will focus on in the section.

To begin with, how to define the wall in the mapping tori? As is discussed in \cite{Cecotti:2010fi}, the wall should be the location where the 4d quiver mutation happens, following the rule:
\begin{align}\label{rulewall}
c: \r_i \rightarrow -\r_i\,, \quad  
T_{\r_i}:\, \r_j \rightarrow \r_j + (\r_i \circ \r_j^+) \r_i \,,
\end{align}
which contains two operations: the flip $c$ is for the BPS state $\r_i$ that should vanish on the wall, and the $T_{\r_i}$ is the Picard-Lefchetz transformation of $\r_i$. It is obvious that the operation $T_{\r_i}$ is just the Dehn twist. Therefore, walls should be at the locations of Dehn twists $T_{\r_i}$ that  are defined as cutting and then gluing along $\r_i$, as is illustrated in \eqref{Tcircle}. In terms of the double branch cover, the wall is:
\begin{align}\bsp
\includegraphics[width=2in]{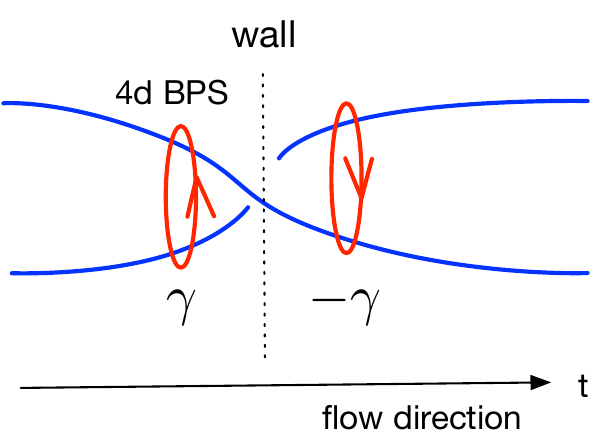} \esp
\end{align}
where the blue lines are actually the branch loci, and braiding describes the Dehn twist; see \cite{Cecotti:2011iy,Cordova:2012xk} for more details on branch covers.

We can then determine the number of walls in the mapping tori. Recall the fact that the monodromy of the mapping tori is the product of Dehn twists in sequence:
\begin{align}\label{monodromycanonical}
M= T_{\gamma_1}T_{\gamma_2}\cdots T_{\gamma_{\mu}} \,.
\end{align}
Each element $\r_i$ corresponds to a mutation and hence a wall, and the number of the walls is the Milnor number $\mu$. These 4d BPS states are distributed on walls, which project to phases on $S^1_t$ in the sequence
\begin{align}
 0<   t_{\r_1} < t_{\r_2} < \cdots t_{\r_\mu} < 2 \pi  \,.
\end{align}

\begin{figure}
    \centering
\includegraphics[width=2in]{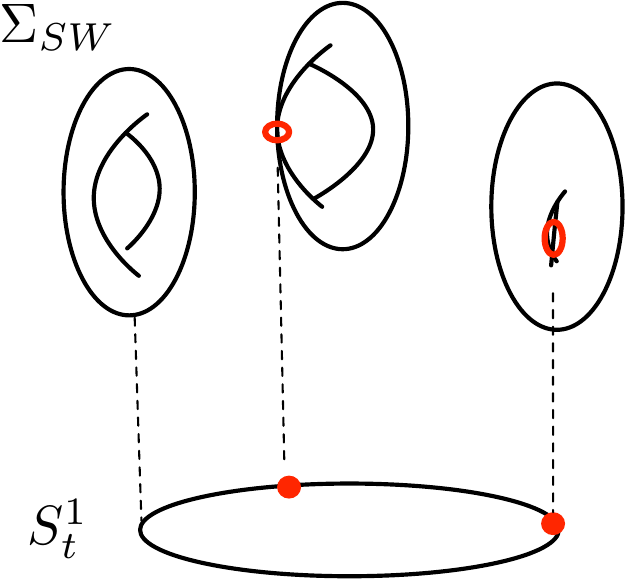}
    \caption{The vanishing circles $\r_i$ shrink on the walls, which project to points on the base. Here, we do not draw the punctures for simplicity.}
    \label{fig:maptori}
\end{figure}

In \cite{Cecotti:2011iy}, it is argued that on each wall there exists only a single 4d BPS state  that should be viewed as a 3d BPS state. This can be interpreted by the RG flow. In the IR limit, the 3d theories become conformal and only massless BPS states could survive, which are just the BPS states undergoing the flip, and vanishing circles on walls give these 3d BPS states, and the rest 4d BPS states on the same page of the mapping tori are massive and hence decouple. Since these 3d BPS states are distributed on different pages of the mapping tori, they cannot link or intersect each other, which is one  feature different from the 4d BPS spectrum. We illustrate the 3d BPS states by the Figure \ref{fig:maptori}. In addition, the number of the 4d BPS states on each page is infinite, as these $\{\r_i$\} could be combined linearly, while this does not happen for the 3d theory.  Thus the 3d BPS spectrum is much more simple than the 4d BPS spectrum.

\subsection{Real central charge}
There is a subtle problem associated with the phases of BPS states. It is know that 3d BPS states have real central charges, and hence do not have any phase, while 4d BPS states have complex central charges and should have phases. The phase is encoded in the central charge by $Z_\r=|Z_\r| e^{i\theta_\r}$. This problem can be solved by the monodromy of the mapping tori.
To begin with, the monodromy changes the complex coordinates by $x'=e^{i[x] \theta}x $ and $y'= e^{i[y]\theta} y$. Then the central charge is also changed 
\begin{align}
Z_\r =\int_{D_\r} dx \wedge dy ~\longrightarrow ~ Z'_{\r} =\int_{D_\r} dx'\wedge dy' =e^{i\theta} Z_{\r} \,.
\end{align}
How to understand this formula?
Assume the $Z_{\r}$ is at the initial point $t=0$ and can have an undetermined phase $\theta_\r$, then after a moment, it goes to $t=\a\theta$, and the central 
charge picks up an additional phase $\theta$. Since $\theta$ arranges from $0$ to $2 \pi$, there will always be a point where $\theta_\r+\theta_\star=0$, and at this point the central charge $Z'_\r$ becomes real $Z'_\r = |Z_\r'|=|Z_\r|$. If this point $\theta_\star$ is the location of the wall, then this 4d BPS state becomes a real 3d BPS state. 

We can keep going, and will meet the next wall. There will be a value $\theta_{\star_2}$ cancels the phase of the next BPS state $\r_2$ attached on the next wall. If we go around the whole base $S^1_t$, all 4d BPS state $\{\r_i\}$ will be reached, and each basic 4d BPS states $\r_i$ is attached on a wall. The 3d BPS states have the central charge  
\begin{align}
Z^{3d}_{\r_i} =|Z^{4d}_{\r_i}| \,.
\end{align}

There is only a relative order for these walls, if  considering the mapping tori as topological manifolds. Now, we can impose the condition that the wall $T_{\r_\star}$ should be located at the place $t_\star =\a \theta_\star$, when projecting it to the base $S^1_t$. This fixes the locations of the walls by the phases of the 4d BPS states. Since the background B-field can be turned on to adjust the phases of 4d BPS states, we cannot go further and determine the values of all phases of 4d BPS states. 

It deserves to mention the relation between the K\"ahler form and the complex coordinates of SW curves, which are related by a proper complex structure
\begin{align}
    \omega_1 =x+\bar{y} \,,\qquad \w_2 =y-\bar{x}
 \,. \end{align}
 Monodromy changes the these coordinates but under control: 
\begin{align}
\w_1' =x'+\bar{y'} = e^{i[x]\theta} ( x + e^{-i\theta} \bar{y}) \,,\quad \w_2' = y'-\bar{x'} =e^{-i[x]\theta} (e^{i\theta}y -\bar{x})  \,,
\end{align}
so the K\"ahler form under the monodromy can be written as
\begin{align}\bsp
    -i k &=d \w_1\wedge d \bar{\w}_1 + d \w_2\wedge d \bar{\w}_2 = 2\, dx \wedge dy-2 \, d \bar{x} \wedge d \bar{y}
    \\
    -ik'&=d \w_1'\wedge d \bar{\w}_1' + d \w_2'\wedge d \bar{\w}_2' = 2\,e^{i\theta} dx \wedge dy-2 \, e^{-i\theta} d \bar{x} \wedge d \bar{y} \,.
    \esp
\end{align}
The 4d BPS central charge $Z_\r' =\int_{D_\r} k' = e^{i\theta} \int_{D_\r} dx\wedge dy  $ is consistent with what we have just discussed.

\subsection{Deformations and moduli space}
We can consider the Coulomb branch deformation of the singularities. 
For the $(A_1,A_{n-1})$ theory, the Seiberg-Witten curve is given by
\begin{align}
    y^2=x^{n} + \sum_{j=2}^{n} u_j x^{n-j} := P_n(x) \,.
\end{align}
The R-charges of coordinates are $ [y] =\frac{ n}{ n+2}$ and  $[x] = \frac{2}{n+2} $, and $[u_j] = \frac{n}{n-j}  $. 
If we change the coordinate $y^2$ to $-y^2$, then this polynomial describes the Milnor fibration of the torus knot $T^{(2,n)}$.

The discriminant is $\Delta= \prod_{1\leq j< k \leq n+1} \left(  e_j -e_k \right)^2  $ where $e_j$ and $e_k$ are the roots of the polynomial $P_n(x)$.
The $\Delta=0$ gives the locations of singularities on the moduli space. The simple singularities locate at $ \{ e_i-e_{i+1} =0, \, i= 1,2, \dots, n-1 \}$ corresponding to vanishing circles $\{ \r_i,  \, i= 1,2, \dots, n-1 \}$. Other higher singularities mean more than one $\r_i$ shrink together.

\begin{figure}[htp!]
        \centering
\includegraphics[width=2.5in]{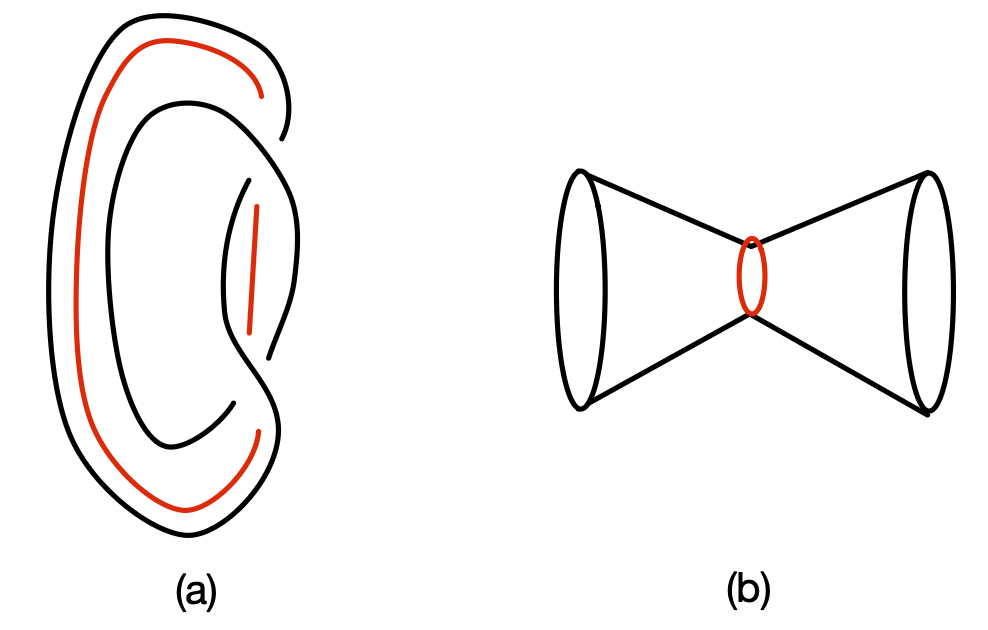}
        \caption{The $(a)$ is the Hopf band for the Seifert surface bounded by the Hopf link. The $(b)$ is equivalent to $(a)$ by only looking at its topology and emphasizing two boundaries. The  parameter $u$ is the length of the red circle denoting $\r$.}
        \label{fig:A1hopf}
    \end{figure}

\vspace{8pt}  \noindent
\textbf{Hopf link.}
When $n=1$, we have the $A_1$ theory, which is a free hypermultiplet with the Seiberg-Witten curve 
\begin{align}
y^2=x^2 +u \,,
\end{align}
which has genus zero, and no Coulomb branch. This Seifert surface has two boundaries forming a Hopf link. Topologically, this curve is a tube and have one vanishing circle. When the  parameter $u=0$, the vanishing circle  shrinks, and the tube becomes two cones connecting at the tip. See Figure \ref{fig:A1hopf} for the  illustration. 

For this theory, the wall crossing only flips $\gamma \rightarrow -\gamma $. 
The mapping tori is $S^3 \setminus \text{Hopf link}$. If filling this complement by 0-surgery, one gets a three-sphere $S^3$.
In addition, 
this theory is the only case that the $S^1_\theta =S^1_t$, as the scale $\a=1$.

\vspace{8pt}  \noindent
\textbf{Trefoil.} This knot characterizes the punctures of the $(A_1,A_2)$ theory.
 Trefoil ${\bf{3}}_1$ is the torus knot $T^{(2,3)}$. 
The deformations of this theory is given by 
\begin{align}
    y^2=x^3+u_2 x+u \,.
\end{align}
The discriminant is given by $\Delta = 4 u_2^3 +27 u^2$ and the solutions are $u_{2,k} = \left(-\frac{27 u^2}{4}  \right)^{\frac{k}{3}}\,, k= 1,2,3 $.
The monodromy  for these three singularities associated with vanishing circles $\gamma_1=(1,0)$, $\r_2=(0,1)$, and anyon $\r_1+\r_2=(1,1)$ , are given by the formula
\begin{align}
T^{(p,q)} =
\begin{bmatrix}
1+pq & p^2 \\
-q^2 & 1-pq
\end{bmatrix} \,,
\end{align}
which gives Dehn twist matrices
\begin{align}\label{T1T2trefoil}
T_{\r_2} =
\begin{bmatrix}
1 & 1 \\
0 & 1
\end{bmatrix} \,,~~
T_{\r_1} =
\begin{bmatrix}
1 & 0 \\
-1 & 1
\end{bmatrix}  \,.
\end{align}
One can check that $ T_{\r_2}(\r_1)=\r_1+\r_2$ and
\begin{align}\label{TTonr1}
   T_{\r_1}\cdot T_{\r_2}(\r_1) =T_{\r_1}(\r_1)+T_{\r_1}(\r_2)  = \r_1+( -\r_1+\r_2 )=\r_2
\end{align}
which is a property \eqref{r1r2r1eqr2} of Dehn twist.
The monodromy of the loop around $\gamma_1$ and $\gamma_2$ is 
\begin{align}\label{A2monodromy}
    M=T_{\r_1} \cdot T_{\r_2} = \begin{bmatrix}
      1 & 1 \\
-1& 0  
    \end{bmatrix}
\end{align}
and has an order $M^6 = 1$, matching with the fact that the order of the mapping tori of trefoil should be $2 \times 3$.
If permute the order of $\r_1$ and $\r_2$, then 
\begin{align}\label{permuteM}
M= T_{\r_1}T_{\r_2}T_{\r_1}^{-1} \cdot T_{\r_1} =T_{\r_1+\r_2} \cdot T_{\r_1} \,.
\end{align}
The vanishing circle $\r_1+\r_2$ is not independent. The monodromy of 4d theory do not picks up all vanishing circles at singularities of the $u$-plane, but only these independent ones, corresponding to the basis of BPS quivers.

\vspace{4mm}\noindent
\textbf{Hurwitz move.}
 In the above \eqref{permuteM}, we have shown there is a conjugation permuting the sequence of Dehn twists in the monodromy.
 The order of these $\gamma$ matters, and only the neighborhood circles $\gamma_i$ and $\gamma_{i+1}$ can be permuted, following the Hurwitz move:
\begin{align}
\gamma_i \triangleright  \gamma_{i+1} := T_{\r_i} T_{\r_{r+1}} T^{-1}_{\r_i}
\end{align}
which is just a quandle action. Therefore, the monodromy is not uniquely represented by Dehn twists and should up to the Hurwitz move.
In terms of the 4d theories, these vanishing circles lead to the 4d BPS quivers. The Hurwitz moves should be interpreted as changing the basis of quivers, and hence be mutations; see also \cite{Xie:2015rpa}.

In general, any diffeomorphism $T_f \in MCG(\Sigma)$ can lead to a conjugation, which acts on $\r$ by 
\begin{align}
T_f \cdot T_{\r} \cdot T_f^{-1} =T_{T_f (\r)} \,,
\end{align}
where $T_f$ means  the Dehn twist along $f$ which is a circle on the Seifert surface and can be expressed as an element in $H_1(\Sigma)$. Then $T_f(\r)$ is a circle obtained by applying $T_{f}$ on $\r$. This conjugation action can be extended to the whole monodromy $M \cdot T_{\gamma} \cdot M^{-1}$. In addition, the conjugation of the monodromy does not change the mapping tori, so there is the equivalence relation $M \simeq T_f \cdot M \cdot T_{f}^{-1}$.
Another property as shown in \eqref{TTonr1} is 
\begin{align}\label{r1r2r1eqr2}
   ( T_{\r_1} \cdot T_{r_2}) (\r_1) = \r_2 \,,
\end{align} 
if these two circles intersect at a point $\r_1 \cdot \r_2^+=1$. 
These relations could derive the braid relation 
\begin{align}
T_{\r_1} \cdot T_{r_2} \cdot T_{\r_1} = T_{\r_2} \cdot T_{\r_1} \cdot T_{\r_2} \,.
\end{align}

\vspace{8pt}\noindent
\textbf{A loop on the moduli space.} \label{loopinmoduli}
The above examples indicate there is a closed relation between the mapping tori and the moduli space of 4d theories.

The Coulomb branch moduli space of the AD theories have singularities at where the circles on the Seiberg-Witten curves shrink. The locations of these singularities are characterized by the discriminant of the Seiberg-Witten polynomial. If consider the bundle of the SW curves fibered over the moduli space, then  the  circles $\r_i$ shrink at singularities. For the $A_2$ theory, we have seen that the Dehn twist $T_{\r_i}$ is equal to the monodromy $M_{\r_i}$ of the singularity. We can draw a loop to pick up the corresponding singularities on the moduli space to match the monodromy $M$ of the Milnor fibration, and then this loop should  cross the branch cuts of these singularities. The total monodromy is just a product $M=M_{\r_1} M_{\r_2} \cdots M_{\r_\u}$. This implies a correspondence between the base $S^1_t$ of the mapping tori and a particular loop $l(\sum_i\r_i)$ on the 4d moduli space. 
Note that there are many other singularities, and we do not pick up them. 
This loop can be illustrated by the following
\begin{align} \bsp
\includegraphics[width=2in]{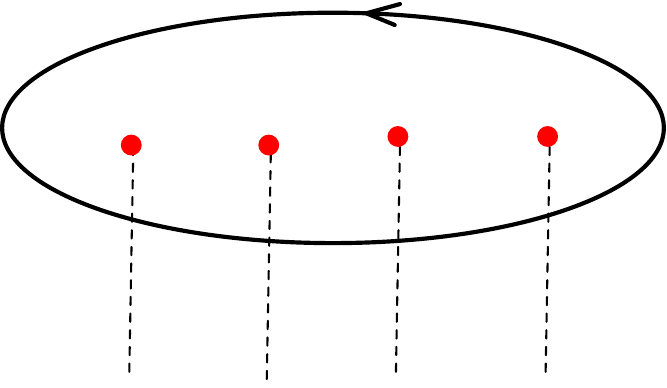}
    \label{loopmoduli}
    \esp
\end{align}
The red dots denote the singularities on the moduli space, and the dash lines denote monodromy cuts. The Hurwitz move is then the permutation of these singularities and monodromy cuts.

There is a simple interpretation for this coincidence. Let us revisit the mapping tori, and analyze the target of the polynomial $f(x,y) \in \mathbb{C}$, which is equivalent to the deformation of the SW curve of the AD theory $x^m+y^n+u=0$ and $u$ parametrizes the complex plane $ \mathbb{C}$. From the definition \eqref{varphi}, the base $S^1_t$ is given by $\varphi =f/|f| = u/|u| =e^{it}$ . Therefore, the base $S^1_t$ should be a loop on the $u$-plane. From this perspective, the Milnor fibration implicitly uses the moduli space of the AD theories as a part of definition.  
Once again, this shows the Seiberg-Witten curves under R-symmetry agrees with the definition of Milnor fibration.

\subsection{Quantum monodromy and chambers}

From the perspective of M-theory, BPS states come from M2-branes. The partition functions of the these M2-branes equal to the open topological string partition functions:
\begin{align}
Z^{\text{open top.}} = Z^{\text{M-theory}} (M_3) = \langle~ \mathcal{T} \prod_\a \mathcal{O}_\alpha(\r_\a)   ~\rangle  \,,
\end{align}
where $\mathcal{T}$ is the time order, and each $\mathcal{O}_\alpha(\r_\a) = \prod_{n=0}^{\inf} (1-q^{n+s+1/2}X_{\r_\a})^{-2s} $ is contributed by a M2-brane ending on the M5-brane.

In \cite{Cecotti:2011iy}, the partition function the of 4d $A_n$-theory is given by the quantum monodromy
\begin{align}
\mathbb{M}^{4d} = \Psi(\r_1)\Psi(\r_1)\cdots \Psi(\r_\u) \cdot \Psi(-\r_1)\Psi(-\r_2) \cdots \Psi(-\r_\u) \,,
\end{align}
which describes the 4d BPS quiver living on the SW curve.
The 4d BPS quiver is
\begin{align}
    \r_1 \rightarrow \r_2 \rightarrow \r_3 \rightarrow \cdots \rightarrow \r_{\mu} \,,
\end{align}
which is defined through the intersections between vanishing circles, and
the arrow means that the Seifert matrix is $S_{ij} =\delta_{i,i+1}$. Each $\Psi(\r_\a) $ is actually the $\mathcal{O_\a}(\r_\a)$, and $X_\a$ satisfy the quantum algebra, since they are Wilson loops.

The 4d theories have a concept called chambers. Each chamber consists a sequence of quivers related by mutations, and the final BPS quiver is 
\begin{align}
    -\r_1 \rightarrow -\r_2 \rightarrow \r_3 \rightarrow \cdots \rightarrow -\r_{\mu} \,,
\end{align}
which is the quiver for the anti-particles  $\{ \r_\a \}$. The particles and anti-particles are CPT conjugated.
There could be many different chambers. The minimal chamber contains the minimal numbers of BPS quivers, and the maximal chamber contains the maximal number of BPS quivers.
For example, for the $(A_1, A_{2})$ theory, the minimal chamber contains three BPS quivers related by two mutations
\begin{align}
    \r_1 \longrightarrow \r_2 ~~\xrightarrow{\r_2}~~ \r_1 \longleftarrow -\r_2 ~~\xrightarrow{\r_1}~~
    -\r_1 \longrightarrow -\r_2 \,,
\end{align}
where the $\r_\a$ texted above the right arrows mean the mutations take place at the vanishing circles $\r_\a$, and the 3d BPS states $\r_\a$ locate on walls.
The maximal chamber is given by firstly mutating the $\r_1$, then   $\r_1+\r_2$ and finally $\r_2$ in sequence, which lead to another sequence of BPS quivers
\begin{align}
    \r_1 \longrightarrow \r_2 ~~\xrightarrow{\r_1}~~ 
    -\r_1 \longleftarrow \r_1+\r_2 ~~\xrightarrow{\r_1+\r_2}~~
    \r_2 \longrightarrow -\r_1-\r_2 
    ~~\xrightarrow{\r_2}~~
    -\r_2 \longleftarrow -\r_1 \,.
\end{align}
The 4d quantum monodromy is invariant under mutations, or in other words, Hurwitz moves. For instance, we have 
\begin{align}
\Psi(\r_1)\Psi(\r_2) = \Psi(-\r_2)\Psi(\r_1) = \Psi(\r_1+\r_2)\Psi(-\r_1) = \cdots \,.
\end{align}

We can define a quantum monodromy for the 3d theories, but it does not seem to be a quantum monodromy. Recall that the 3d BPS states only locate on the walls where mutations could happen, so the open topological string partition functions for the $(A_1,A_2)$ theory is 
\begin{equation} \label{trefoidual}
Z({S^3\setminus \textbf{3}_1}) =
\langle \mathcal{O}_{\r_2} \mathcal{O}_{\r_1} \rangle =
\langle \mathcal{O}_{\r_1} \mathcal{O}_{\r_1+\r_2} \mathcal{O}_{\r_2} \rangle  \,,
\end{equation}
which is the SQED-XYZ duality discussed in \cite{Cecotti:2011iy,Dimofte:2011ju}. The 3d partition function should be equivalent for different chambers. In the minimal chamber, there are two walls on the mapping tori $S^3\setminus \textbf{3}_1$, while on the maximal chamber, there are three walls. However, in the XYZ theory, there is  a cubic superpotential $\W= \phi_{\r_1} \phi_{\r_2}\phi_{\r_1+\r_2}$ coupling three chiral multiplets and hence cancels the contribution from the additional BPS state. Thus this duality is a Seiberg duality. 

For more generic theory, such as $(A_1, A_n)$ theory,  if the mutations happen on the mapping tori are $\gamma_1, \gamma_2,\dots, \gamma_\mu$ in sequence, it returns the minimal chamber. The 3d theory have the partition function 
\begin{align}\label{3dAnquantummonodromy}
    Z(S^3 \setminus T^{(2, n+1)}) =\langle \mathcal{O}_{\r_1} \mathcal{O}_{\r_2}\cdots  \mathcal{O}_{\r_\mu} \rangle \,.
\end{align}
One can also analyze dualities for this theory, related by non-minimal chambers, which is complicated, but the basic duality is still caused by the cubic superpotential and the local Seiberg duality, analogous to the case of $\textbf{3}_1$ knot. 

In short,
at first sight, there is no relation between the monodromy and quantum monodromy. The former is  motivated by the topology and geometry, and the latter is motivated by quantum algebra. But their relation becomes obvious when it comes to 3d theory on mapping tori: each Dehn twist $T_{\r_\a}$ is replaced by an operator $\mathcal{O}_\a(\r_\a)$.

However, \eqref{3dAnquantummonodromy}
still cannot reflect how the M2-branes are charged under the gauge groups. In the next section, we will show that 
using the Dehn surgery, one can determine the locations of  these M2-branes in the three-manifolds and how the boundaries of the M2-branes link to the torus knot $K$. 

\vspace{4mm}
\noindent
\textbf{The full 3d BPS spectrum.}
 To get the full partition function, the 4d  anti-BPS spectrum should be encoded in the 3d partition function. However, at first sight the monodromy $M=T_{\r_1} T_{\r_2}\cdots T_{\r_u}$ does not reflect how the anti-BPS particles $\{ -\r_\a\}$ are included in the quantum monodromy 
\begin{align}
\mathbb{M}^{4d} = \Psi(\r_1)\Psi(\r_1)\cdots \Psi(\r_\u) \cdot \Psi(-\r_1)\Psi(-\r_2) \cdots \Psi(-\r_\u) \,.
\end{align}
An idea can be used to solve this puzzle. Recall that in the minimal chamber, after a series of mutations from right to left, the BPS quiver starts from $ \r_1 \rightarrow \r_2 \rightarrow \cdots \rightarrow \r_\u$ and finally becomes $ -\r_1 \rightarrow -\r_2 \rightarrow \cdots \rightarrow -\r_\u$, which, if we apply the same series of mutations again, it recovers the original BPS quiver:
\begin{align}
-\r_1 \rightarrow -\r_2 \rightarrow \cdots \rightarrow -\r_\u \quad  \cdots \longrightarrow \cdots \quad \r_1 \rightarrow \r_2 \rightarrow \cdots \rightarrow \r_\u \,.
\end{align}
This implies that the minimal chamber can be doubled. 

For the mapping tori, if we let the 4d BPS quiver goes around the base $S^1_t$  one time, BPS quiver becomes its negative, and if we continue and go around the base one more time, BPS quiver recovers itself.  This property is caused by the fact that when mutation is applied, the vanishing circle is flipped from $\r_i$ to $-\r_i$, so we need to mutate two times to produce the full spectrum. Then the full 3d partition function should be 
\begin{align}
Z^{3d} (S^3 \setminus K)=  \Psi(\r_1) \Psi(\r_2) \cdots \Psi(\r_\u) \cdot \Psi(-\r_1) \Psi(-\r_2) \cdots \Psi(-\r_\u) \,.
\end{align}
The non-commutative relation between $\Psi(\r_\a)$ is ignored, because in the minimal chamber these 3d BPS states live on different pages of the Milnor fibration, and they never really meet. In other chambers, the cubic superpotential can be introduced to change the order of $\Psi(\r_\a)$. The significance of this formula is that  the 3d partition function captures the full 4d quantum monodromy, although the 4d quantum relation for $X_\r$ is deleted.

It is possible to see this from the M2-M5 brane configuration. On each wall of the mutation locates a M2-brane, which is oriented, and we can assume it is pointing up (the $\r_\a$). This M2-brane disc reverses its orientation when crossing the wall and becomes pointing down (the negative $-\r_\a$). After all mutations of the minimal chamber, all M2-branes point down. Now, we go around the base circle $S^1_t$ again. When crossing the first wall this time, the M2-branes become pointing up again, namely $-\r_1 \rightarrow \r_1$, and so on. When all walls are crossed, all M2-branes point up again. Therefore, from this analysis, we can see on each wall, both BPS particle $ \r_\a$ and its anti-particles $-\r_\a$ are captured. In addition, if we do not consider gauging the global symmetry coming from the boundary of the mapping tori of torus knots, the full 3d partition functions should be equal to the 4d quantum monodromy.

However, this interpretation becomes tricky for non-minimal chambers, but this is not a problem, because non-minimal chambers can be obtained from minimal chambers by introducing superpotentials and additional BPS particles. We will go back to this in the next section. 

\vspace{4mm} \noindent
\textbf{N-fold cover.}
$N$-fold cover of the mapping tori of fibered knots is defined as 
\begin{align}
{M_h^N} = \frac{\Sigma \times [0,1]}{ (x,0) \sim (h^N(x), 1)} \,,
\end{align}
which is given by gluing $N$ copies of the mapping tori $M_h=\Sigma \times_h S^1$.  Reversely, its  orbifold is the mapping tori itself: $M_h^N / \mathbb{Z}_N = M_h$. 
The 4d monodromy of the $N$-fold cover is given by 
$
  \mathbb{M}= \mathbb{M}^N
$. 

There is a subtle point. This $N$-fold  is indeed a mapping tori, but generically is not a mapping tori of  any fibered knot, so this $M_h^N$ may not be represented as the Milnor fibration. But still since the $M_h^N$ has a monodromy of finite order, it can be represented as a Seifert manifold fibered over a sphere $S^2$.

\section{Surgeries}
\label{sec:surgery}
The mapping tori of torus knots can be viewed as the knot complements. This nice property makes the analysis much more easier than other three-manifolds. In this section, we switch the language to Dehn surgery and see how the BPS states are distributed in the knot complements.

\vspace{8pt}\noindent
\textbf{Dehn surgery.}
Lickorish-Wallace theorem tells that any closed three-manifolds can be given by surgeries over links, and each component of the link can be an unknot. This means that any three-manifolds given by $S_f^3(K)$ can be resolved into links of unknots:
\begin{equation}
K ~\xrightarrow{\text{Kirby moves}} ~ \bigcirc_{k_1} \cup  \bigcirc_{k_2} \cup  \cdots \cup \bigcirc_{k_n}\,,
\end{equation}
which in many cases leads to simplification, and the knot $K$ is called surgical knot.
Each $\bigcirc_k$ should be applied a Dehn surgery by filling in a solid torus with the gluing map $T^k \in SL(2,\Z)$. For instance, the three-sphere is $S^3=S^3(\bigcirc_{\pm 1})$. For a given closed three-manifold, there are infinite many equivalent surgeries related by Kirby moves and Rolfsen twists. 

The surgery representation of three-manifold has a nice physical interpretation in gauge theories. If the three-manifold is wrapped by a single M5-brane, then each $\bigcirc_k$ corresponds to a gauge group $U(1)_k$ with Chern-Simons level $k$. The linking number $K_{ij} = \bigcirc_{i-th} \circ \bigcirc_{j-th}$ is the effective mixed Chern-Simons level between the gauge group pair $U(1)_{i-th} \times U(1)_{j-th}$. In addition, the Kirby moves are dualities of 3d gauge theories \cite{Gadde:2013sca}. 

In the presence of the M2-brane, the boundary of the M2-brane is a circle $\r$  that should link to  surgical knots, and the linking number between $\r \circ K $ is the charge of the chiral multiplet \cite{Cheng:2023zai}. Even in the presence of M2-branes, the Kirby moves of both blow-up/down and handle sliding of surgical knots are still preserved \cite{Cheng:2024ybd}. However, we still do not clearly know how to perform Kirby moves directly on $\r$ coming M2-branes. Naive application of Kirby moves on $\r$ often leads to mistakes.

A subtle issue is on the plumbed graphs that are often used to denote the surgical links and Kirby moves. Kirby moves  often transform some parts of the surgical links into knots, so in many cases it is more better to just draw  links and knots to avoid potential mistakes.

\subsection{3d theories}
Using the Kirby moves, we can simplify the band graphs of  torus knots, and in principle write down the associated abelian 3d theories. 

\vspace{8pt} \noindent
\textbf{$A_1$ theory.}
The most simple example is the $A_1$-theory as shown in Figure \ref{fig:A1hopf}. If filling the boundary by $0$-surgery, we get a 3d bifundamental chiral multiplet with charges $(+1,-1)$, denoted by the plumbed graph
\begin{align} \label{A1plumb}
\bsp
\includegraphics[width=1in]{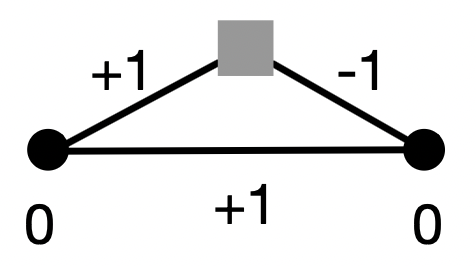}
\esp
\end{align}
This plumbed graph can be transformed by Kirby moving  the gauge groups (black node) and the chiral multiplet (gray box represent $\r$) to get $\bullet_{-1}-\textcolor{gray}{\blacksquare}$ which is dual to a free chiral multiplet, and the three-manifold is actually the three-sphere $S^3_{-1}$. In addition,
Kirby moves could transform the Hopf band to the following link by removing the twisted region of the Hopf link
\begin{align} 
\bsp
\includegraphics[width=1.5in]{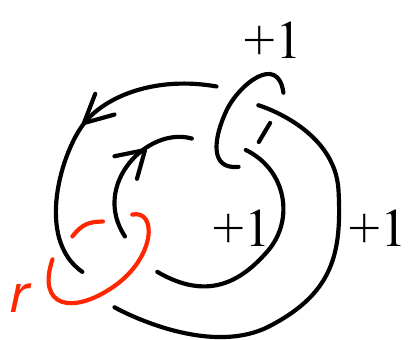}
\esp
\end{align}
which has the gauge group $U(1)\times U(1)\times U(1)$.

\vspace{8pt} \noindent
\textbf{$(A_1,A_2)$ theory.}
The right-handed trefoil is the fibered knot for this theory. We start from its band graph, on which vanishing circles $\r_1$ and $\r_2$ intersect at one point. After the Reidemeister moves, the band graph is reorganized, and we separate the $\r_1$ and $\r_2$, since $\r_1$ and $\r_2$ locate at two different pages of the mapping tori. Finally, we deform the graph to the standard trefoil, and show the locations of $\r_\a$. We illustrate this process in the following:
\begin{align} 
\includegraphics[width=5.5in]{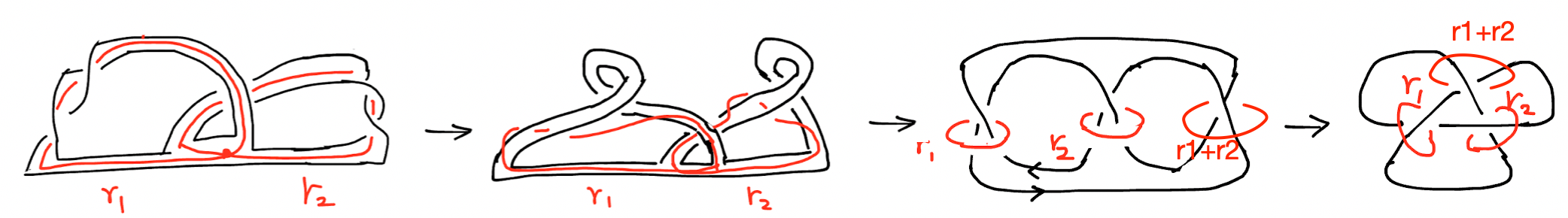}
\end{align}
In the first two graphs, we consider a page of the mapping tori, and hence $\r_1$ and $\r_2$ are on the same Seifert surface.
In the last two graphs, we consider the mapping tori as the knot complement. The additional circle $\r_1+\r_2$ as the connected sum of $\r_1$ and $\r_2$ should appear at the third cross. This circle is not in the basis of the 4d BPS quiver $\r_1 \rightarrow \r_2$, but can be added by hand to keep the symmetric beauty of the knot. Adding this vanishing circle is equivalent to going to the maximal chamber, as shown in  the duality \eqref{trefoidual}.
In addition, the cubic superpotential 
\begin{align}
W=\phi_{\r_1}\phi_{\r_1+\r_2}\phi_{\r_2}
\end{align}
should exist, and 
corresponds to the triangle region bounded by three lines in the trefoil, and on each tip of the triangle locates a BPS state.

It can be computed that if using the standard definition of the linking number, the linking number between each $\r_\a$ and the trefoil is zero,  but actually $\r_\a$ and the trefoil are linked and cannot be separated. This implies that when using plumbed graphs to study equivalent surgeries, one needs to be careful. One can try to use Kirby moves to transform the trefoil into surgical links, just like what we did for the $A_1$ theory, and read off the gauge groups and charges for chiral multiplets, but it seems a bit tricky. Fortunately, this problem can be bypassed by considering the $(A_1,A_3)$ theory.

\vspace{4mm}\noindent
\textbf{$(A_1,A_3)$ theory.}
The associated torus knot  for this theory is $T^{(2,4)}$, and the band graph and locations of $\r_\a$ are shown in the following:
\begin{align} 
\bsp
\includegraphics[width=5in]{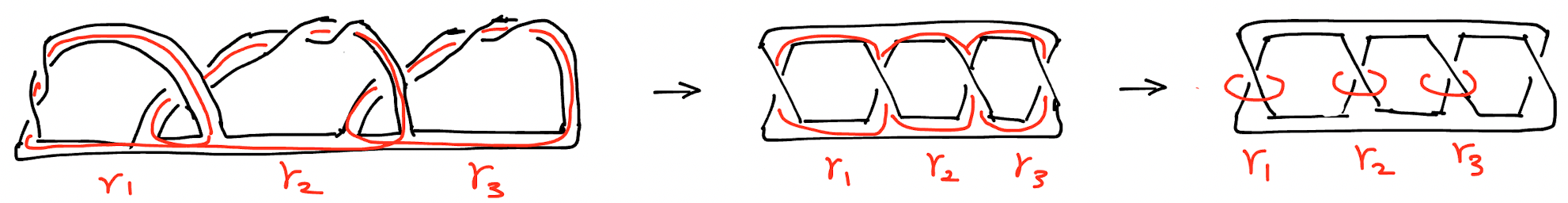}
\esp
\end{align}
The 4d BPS quiver is $\r_1 \rightarrow \r_2 \rightarrow \r_3$.
In this case, the last cross on the right graph can be assigned with a state $\r_1+\r_2+\r_3$ such that a superpotential $W=\phi_{\r_1}\phi_{\r_2}\phi_{\r_3}\phi_{\r_1+\r_2+\r_3}$ is turned on, corresponding to the shaded rectangle of the following graph 
\begin{align} \label{A1A3figure}
\bsp
\includegraphics[width=1.8in]{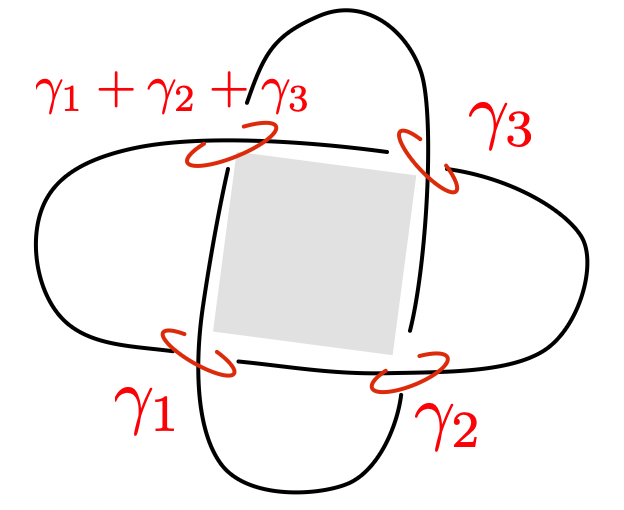}
\esp
\end{align}
Notice that this mapping tori have two boundaries, and a flavor symmetry for the $(A_1,A_3)$ theory. If filling these two boundaries by solid tori, the 3d theory has two gauge groups $U(1) \times U(1)$ with mixed Chern-Simons levels. The 3d BPS states are bifundental chiral multiplets with charges $(+1,-1)$. For this theory, we do not meet the vanishing linking number issue that appears in $(A_1, A_{2n})$ theory. 

To relate this theory to $(A_1,A_2)$ theory, one can first change the orientation of one surgical unknot of the torus knot $T^{(2,4)}$, which flips the signs of the related linking numbers, and then we trivialize one BPS circle such as $\r_3$ and replacr it by a surgical unknot $\bigcirc_0$ with zero framing number, which means the M2-brane of $\r_3$ is ignored and hence $\Psi(\r_3) \rightarrow 1$. Using the fact that the $\bigcirc_0$ is equivalent to performing connected sum, the surgical link $T^{(2,4)} \cap \bigcirc_0$ is equivalent to $T^{(2,3)}$, and in this way the $(A_1,A_3)$ theory reduces to $(A_1,A_2)$ theory. This reduction procedure works also on the plumbed graphs consisting of black nodes and gray boxes, such as \eqref{A1plumb}, and can be used to read off the 3d theory for $(A_1,A_2)$ theory. This is how we bypass the vanishing linking number issue.

Note that these 3d theories arsing from $(A_1, A_n)$ theories are in the minimal chamber, if taking into account that the superpotential cancels the additional state $\sum_{\a}\r_\a$ which is the sum of other independent 3d BPS states.
In the case of $(A_1,A_2)$ theory, the maximal chamber is caused by adding the cubic superpotential. For $(A_1,A_3)$ theory, other chambers are complicated, and there could be many 3d BPS states in non-minimal chambers. Different chambers give the same partition function, and hence chambers just mean dualities.
However, different chambers could assign different numbers of BPS states to the mapping tori.
If this number is larger than Milnor number, then there are redundant 3d BPS states and associated superpotentials.

Let us explain this by the 3d BPS spectrum. Let us start from a chamber of $(A_1,A_3)$ theory, which contains a sub-chamber:
\begin{align}
&\r_1 \rightarrow \r_2 \rightarrow \r_3  ~~\xrightarrow{\r_1}~~ -\r_1 \leftarrow \r_1+\r_2 \rightarrow \r_3 ~~\xrightarrow{{\r_1+\r_2}}~~ 
\\ \nonumber
&\r_2  \longrightarrow
-\r_1-\r_2 \longleftarrow \r_1+\r_2+\r_3 ~~\xrightarrow{\r_1+\r_2+\r_3} ~~ -\r_2 \leftarrow \r_2+\r_3 \rightarrow -\r_1-\r_2-\r_3 \\ \nonumber 
& \xrightarrow{\r_2+\r_3} \r_3 \rightarrow -\r_2-\r_3 \leftarrow {-\r_1} ~~\xrightarrow{\r_3}~~ -\r_3 \leftarrow-\r_2 \leftarrow -\r_1   \,.
\end{align}
The 3d BPS states on the walls are $\{ \r_1, \r_1+\r_2,  \r_1+\r_2+\r_3, \r_2,\r_2+\r_3, \r_3\}$. This chamber contains two sub-chambers, and we do not show the other one for simplicity, but in total the 3d BPS spectrum should capture all 4d BPS states of this chamber.

From above two examples, one can observe that the mapping tori of torus knots give a parallel construction to the DGG construction \cite{Dimofte:2011ju}, since the three-manifolds we are considering are non-hyperbolic, and the DGG construction excludes non-hyperbolic manifolds. In the DGG construction, the basic building block is the tetrahedron and each is associated with a chiral multiplet, while in our case the basic thing is the Hopf band, on which the chiral multiplet is the vanishing circle, coming from the 4d BPS state that survives on the wall in the deep IR. 

\vspace{5mm}\noindent
\textbf{Superpotentials.}
The above examples show the polynomial superpotential is encoded in the relation between BPS states. If a BPS sate is a linear combination of some other BPS states, then a superpotential is indicated; see also \cite{Cheng:2024ybd}. 
 In practical, to see if there is a superpotential, we can move a group of $\r_\a$ and let them go around the surgical knot, and see if these $\r_\a$ can move to the same location and form a connected sum $\sum_\a \r_\a=0$ such that superpotential can be generated. Therefore, the polynomial superpotential is encoded in the topological structure of the knot itself. The trefoil and $(2,4)$-torus knot are good examples. As for superpotentials caused by mutations, we can consider a local graph of the $(A_1,A_3)$ theory in \eqref{A1A3figure}. For instance, if $\r_2$ and $\r_3$ are exchanged under the Hurwitz move, a cubic superpotential is generated and the additional state $\r_2+\r_3$ should be introduced, as shown in the following:
\begin{align}
    \bsp \includegraphics[width=1.5in]{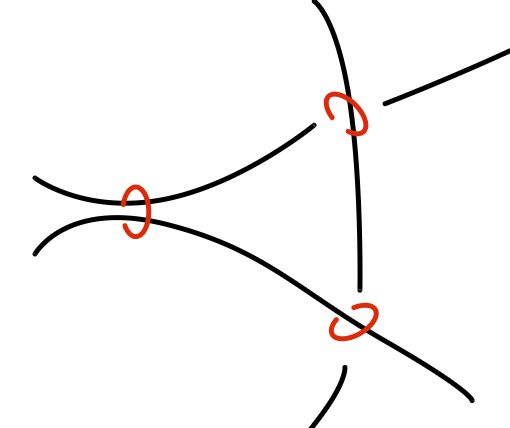} 
    \esp
\end{align}
Repeating this process, other superpotentials can be obtained, and hence the whole non-minimal chambers. This shows again that 4d chambers descend to 3d dualities.

The monopole potentials are much more confusing, but can be engineered by M2-branes. In \cite{Cecotti:2010fi}, the monopole potentials are discussed, which are given by M2-branes wrapping on two-spheres  in three-manifolds, and may couple to 3d BPS states $\r_\a$. The rest dimension of the monopole M2-branes extends into the four-manifolds bounding the three-manifolds. To find these two spheres, recall that on some locations (walls) of the mapping tori there are shrinking circles $\r_\a$, which sweeps out cigars during the R-flow. Sometimes, these cigars could join and merge into spheres in the bundle. An example is the mapping tori of trefoil, where the cigars of $\r_2$, $\r_1+\r_2$, and $\r_1$ merge into a sphere. We illustrate this in the following figure:
\begin{align}\label{monopole}
\bsp
\includegraphics[width=3in]{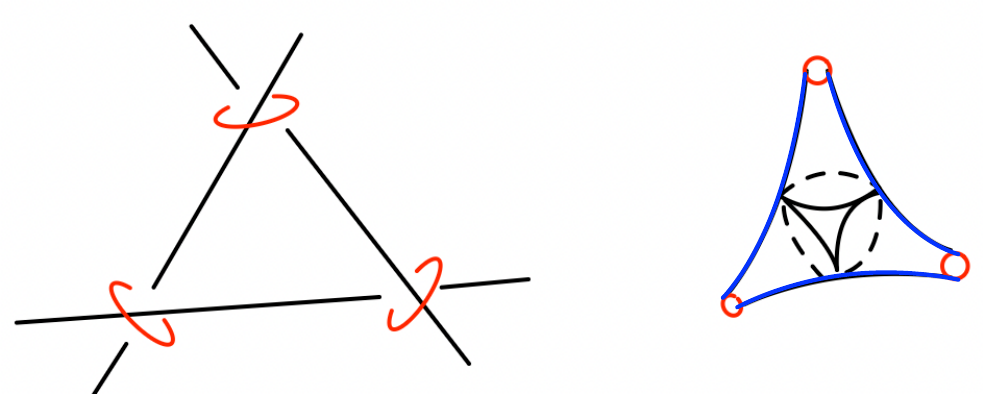}
\esp
\end{align}
This configuration is actually the same as in \cite{Cecotti:2011iy}, and can be described more precisely by the flow of the branched loci (the blue lines in \eqref{monopole}). Branched loci also form a knot, and is related to the fibered knot, but  more effort is needed to discuss these branched knots, and we will discuss more in future work.

Another aspect of monopole potentials is related to Kirby moves of surgical knots, which change the number of gauge groups, and hence change the number of topological symmetries. In order to realize the Kirby moves as dualities, the topological symmetries must be preserved, and  monopole potentials should be turned on to break additional topological symmetries.

\vspace{4mm}
\noindent
\textbf{Flavor symmetries.}
For the $(A_1,A_{2n+1})$-theory, the corresponding mapping torus has two boundaries forming a link, while for the $(A_1, A_{2n})$-theory, there is only one boundary as the irregular puncture. For the 4d $(A_1,A_{2n+1})$-theory, there is a flavor symmetry with a mass parameter taking value at the residue of the puncture, which could be lifted as a flavor symmetry of the corresponding 3d theories. 

For the knot complement $S^3\setminus K $, the holonomy along the meridian of the boundary is not vanishing, while the holonomy along the longitude of $K$ vanishes, so the meridian of $K$ should be served as a topological symmetry, even if the boundary is filled by a solid torus through 0-surgery.

\section{Enhancement}\label{sec:enhancement}
In this section, we notice that the mapping tori of torus knots have a very good structure called transverse holomorphic foliation, which allows a partial topological twisting that leads to  $\N=4$ supersymmetry for 3d gauge theories. The conclusion is consistent with \cite{Assel_2023}.

\subsection{Transverse holomorphic foliations}

The 3d theories labeled by three-manifolds often have a supersymmetry of $\N=2$. In particular cases, we expected more supersymmetries. If the monodromy of the mapping tori is trivial, 3d theories coming from 4d $\N=2$ theories by a circle reduction have $\N=4$ \cite{Benini:2010uu}. 
To preserve supersymmetries, we should
ensure the Killing Spinor equations (KSE) have solutions on  three-manifolds. If the three-manifolds have transverse holomorphic foliations (THF), as discussed in \cite{Closset_2013,Closset_2014}, then more supersymmetries are allowed. 

Let us review how the topological twist is applied to gauge theories.
The topological twist  uses the $SO(5)_R$ R-symmetry of the 6d $(2,0)$ theory. When the 6d $(2,0)$ gauge theories is compactified on the Seiberg-Witten curve, the R-symmetry breaks $SO(5)_R \rightarrow SO(2)_R \times SO(3)_R$ where the $SO(2)_R$ is used to twist with the rotation symmetry of $SO(2)_\Sigma$ on the SW curve: $SO(3)_R\times SO(2)_\Sigma \rightarrow U(1)_r$. After the compactification, the $U(1)_r \times SO(3)_R$ are the R-symmetry for the 4d theories. If compactifying on the three-manifold $M_3$, the 6d R-symmetry splits as $SO(5)_R \rightarrow SO(3)_R \times SO(2)_R$, where the $SO(2)_R$ is left as the R-symmetry of the resulting 3d $\N=2$ theory, and the $SO(3)_R$ is used for topologically twisting with the rotation symmetry $SO(3)_{M_3}$  of the three-manifold to get $\N=2$, which break $3/4$ of supersymmetries of the 6d theories. 

The above is the standard topological twisting. The partial topological twisting is defined as follows:
The 6d (2,0) theories live on M5-branes, and the M5-branes wrapping on Seiberg-Witten curves (Milnor fiber) lead to 4d $\N=2$ theories. The standard topological twist is applied on Seiberg-Witten curves. To extend this twist to mapping tori, we should consider the monodromy. Depending on how many solutions of the KSE on SW curves are invariant under the monodromy $h$ of the mapping tori $\Sigma_{SW} \times_h S^1_t$, we can determine the number of the supersymmetries for the 3d theories. Fortunately, since the mapping tori of torus knots have THF, all solutions are preserved and hence the 3d theories have $\N=4$. Note that the $U(1)_r$ part of the R-symmetry should not be  called topologically twisted with $S^1_t$ of the mapping tori but R-twisted with it. 

In other words, we only need do the topological twisting on each page (Seiberg-Witten curve) of the mapping tori. The spinors on each page are constant and do not depend on the R-twisted circle $S^1_t$, and thus are constant on the whole mapping tori. This partial topological twisting  break a half of the supersymmetry  of 6d theories and leads to 3d $\N=4$ theories.

In general, three-manifolds allow standard topological twisting on the whole three-manifold, and have  $\N=2$, while mapping tori with THF have  partial topological twisting leading to $\N=4$. Then the question is that could these two types of topological twisting lead to equivalent theories? We think the answer is yes. The 3d theories look like $\N=2$, since they have Chern-Simons levels and superpotentials, which are features of $\N=2$ theories, but actually have $\N=4$ in disguise.

\vspace{8pt} \noindent
\textbf{Compare with \cite{Assel_2023}.} This paper discussed the enhancement of 3d theories to $\N=4$, obtained by assigning $T_N$-theories to particular Seifert manifolds $M(0; 1/k_1,1/k_2,1/k_3)$, satisfying the enhancement condition
\begin{align}\label{enhancementcondition}
1/{k_1}+1/{k_2}+1/{k_3} =0 \,.
\end{align}
Although the analysis is on non-abelian $T_N$ theories,  the conclusion is driven from the superpotential, and hence conclusion also applies to abelian theories. An example is the Seifert manifold $M(0, 1/2,-1/3,-1/6)$, which is nothing but the three-manifold $S^3_0(\mathbf{3_1})$. This allows us to see a connection with our construction.

Let us first review the $t/u$-surgery on the $(p,q)$-torus complements, which give the Seifert fibered manifolds that are circle bundles over a two sphere $S^2$:
\begin{align}
S^3_{t/u}\left(T^{(p,q)}\right)=    \left(S^2,  (p,r),(q,s),(t-pqu,u)  \right) \,,~~ \text{if}~ t/u \neq pq \,,
\end{align}
where $(r,s)$ is any pair with $ps+qr=1$; see \cite{Moser1971,Martelli2016}.
For the 0-surgery, a nice property emerges: Euler number vanishes
\begin{align}\label{eulerzero}
 \frac{r}{p}+ \frac{s}{q} -\frac{1}{pq} =0 \,.
\end{align}
This manifold has an infinite cyclic $\pi_1(M_3) =\mathbb{Z}$. 
In addition, these Seifert manifolds have a finite cover $S^2\times S^1$. These Seifert manifolds can be represented by double branched covers over Montesinos knots as  branched loci.

After changing notation,  Seifert manifold $S_0^3\left(T^{(p,q)}\right)$ is just $M\left(0, \frac{r}{p}, \frac{s}{q}, -\frac{1}{pq} \right)$. 
Thus, the enhancement condition \eqref{enhancementcondition} just means Euler number is zero \eqref{eulerzero}, which is a property of the mapping tori of torus knots $S_0^3\left(T^{(p,q)}\right)$ themselves. This indicates that the 3d $\N=4$ theories obtained in \cite{Assel_2023} and our construction should be the same, at least for abelian theories.

In addition, in the last section, we mentioned that monopole potential is given by M2-branes wrapping on a two-sphere in the three-manifold. Since the Seifert manifolds are circle bundles over a two-sphere and the Euler number vanishes, the base sphere can be lifted to a sphere in the circle bundle, which is an essential sphere in the mapping tori, and hence carries a M2-brane that generates a monopole potential, as shown in \eqref{monopole}.

\subsection{Rank zero theories}
Let us consider another class of 3d theories called rank zero theories \cite{GangYamazaki2018}, which are believed to be obtained by a circle reduction of 4d AD theories and show a $\N=4$ enhancement. This relation between 4d and 3d is somehow confirmed by 4d quantum monodromy and partition functions. 

We expect that there is no much difference between the two step reduction $\Sigma_{SW} \times S^1$ and the one step reduction of mapping tori $\Sigma_{SW} \times_h S^1$, so we conjecture our construction should relate to the 3d rank zero theories. The only thing we should add here is that the trivial production $\Sigma_{SW} \times S^1$ should be lifted to the mapping tori of torus knots $\Sigma_{SW}\times_h S^1_t$ with non-trivial monodromies.

Since 3d partition functions capture all 4d BPS states, 3d partition functions are identified with the 4d quantum monodromy as we discussed in the last section, and if proper parameters are chosen, equal to the Schur indices of AD theories.
The 4d quantum monodromy, if applied mutations and some identities to cancel some chiral multiplets,  will take the form of the vortex partition functions (half indices) of certain 3d $\N=2$ theories that are abelian theories with mixed Chern-Simons levels, with monopole superpotentials governing the topological symmetries. Following this logic, many AD theories lead to the 3d ranks zero theories; see e.g.\cite{GaiottoKim2025}. 

In \cite{Gang:2023rei,Dedushenko:2023cvd}, a class of rank zero theories are obtained.
These rank zero theories are simple:
\begin{align}\label{GKStheory}
U(1)^r_{K_r} + \sum_{i=1}^r \Phi_{i} \end{align}  where $\Phi_i$ is the fundamental chiral multiplet charged only under the $i$-th gauge group, and effective mixed Chern-Simons levels are
\begin{align}\label{GangK}
  K_r = 2 \begin{pmatrix}
1 & 1 & 1 & \dots & 1 & 1 \\
1 & 2 & 2 & \dots & 2 & 2 \\
1 & 2 & 3 & \dots & 3 & 3 \\
\vdots & \vdots & \vdots & \ddots & \vdots & \vdots \\
1 & 2 & 3 & \dots & r-1 & r-1 \\
1 & 2 & 3 & \dots & r-1 & r
\end{pmatrix} \,.
\end{align}
 We do not copy the monopole superpotentials to here for simplicity. 
The vortex partition functions can be written down with these information.

Since the 4d quantum monodromy is equal to the Schur index \cite{CordovaShao2016}, one can directly use Schur index.
In \cite{FodaZhu2020}, the Shur index is simplified to the form 
\begin{align}\label{shurexample}
I^{(A_1,A_{2r})} = \sum_{n_1 n_2,\cdots, n_r =0 }^{\inf} \frac{ (-\sqrt{q})^{ 2 n_1^2 +2 n_2^2 + \cdots + 2n_r^2  }   q^{ n_1+n_2 + \cdots + n_r} }{  (q)_{n_1-n_2}(q)_{n_2-n_3} \cdots (q)_{n_{r-1}-n_r} (q)_{n_r}   } \,.
\end{align}
If compare this formula with the vortex partition functions from localization computation, one can identify that each $n_i$ is a vortex number associated to a gauge group, and each $(q)_{n_i-n_j}$ is for a bifundamental chiral multiplet. The mixed Chern-Simons levels can be read from the term $(-\sqrt{q})^{K_{ij}n_in_j}$ and hence $K_{A_{2r}},_{ij}= 2 \delta_{ij}$.

Recall that 
4d quantum monodromy contains $4r$ number of $\Psi_{\r_i}$, which can be reduced to $r$ chiral multiplets through various identities. Then how to relate this Schur index \eqref{shurexample} to the theories labeled by $K_r$ in \eqref{GangK}? We notice that the Kirby moves of the three-manifold can be used. If applying the handle slides (the second type of Kirby moves, which are represented as the linear recombinations of $n_i$),
the vortex partition function of \eqref{GKStheory}
can be written in the form \eqref{shurexample}. Namely,
\begin{align}
    K_r ~~\xrightarrow{ \text{handle slides} }~~ K_{A_{2r}} \,.
\end{align}
These handle slides on gauge groups are discussed in \cite{Cheng:2024ybd}, and could represent identities simply by handle sliding surgical knots and links.

The most simple rank zero theory is the case $r=1$, which is $U(1)_2+1 \Phi$ and the chiral multiplet is charged one in fundamental representation. This theory is supposed to be obtained from the 4d $(A_1,A_2)$-theory. We conjecture this theory descends from the 3d theory labeled by $S^3_0(\textbf{3}_1)$.

\section{Discussion}

In this note, we discussed the mapping tori and the  application to 3d gauge theories. There are some open problems:
\begin{itemize}
\item The $N$-fold cover is also useful. The $N$-fold cover can have the potential to engineer non-abelian theory by putting $N$ M5-branes on the mapping tori $S_0^3(K)$, and then build the connection with the $\hat{Z}$-invariants proposed in \cite{Gukov:2016gkn,Gukov:2017kmk}.

\item In this note we only discussed the torus knots. The AD theories of other types \cite{Akbulut_2001} can also be discussed.

\item One can consider all fibered knots through Murasugi sum of band graphs, following the construction \cite{Gabai1983,Stallings1978}.

\item The Giroux correspondence may be used to build the relations between the contact structures, Legendrian knots, three-manifolds, and gauge theories. 
\item Mapping tori are boundary of four-manifolds, so we expect many structures can be extended to four-manifolds and 2d gauge theories.

\item Any Seifert manifolds can be represented by double branched covers, so the mapping tori should be included in this more broad construction. This is a work to appear soon. 

\item One can connect mapping tori to the 3d/3d correspondence  \cite{Dimofte:2011ju,Dimofte:2013lba,TerashimaYamazaki2011}, and relate tetrahedron decomposition to mapping tori.

\item It is interesting to see if mapping tori can understand the 4d mirror symmetry proposed in \cite{Shan:2023xtw}.

\end{itemize}

\acknowledgments
I would like to thank Mykola Dedushenko and especially Hao Zou for helpful discussions. This work is supported by the start-up grant of SIMIS.

\appendix

\bibliographystyle{JHEP}

\bibliography{biblio.bib}

\end{document}